\documentclass[runningheads]{llncs}
\usepackage{graphicx}
\usepackage[indentfirst=false,leftmargin=10pt,rightmargin=10pt]{quoting}
\begin{document}
\title{Developing an Augmented Reality Tourism App through User-Centred Design\\ (Extended Version)\thanks{This paper is an expansion of our earlier publication Williams, Yao \& Nurse, 2017, ToARist: An Augmented Reality Tourism App created through User-Centred Design, 31st British Human Computer Interaction Conference (BHCI)}}
%
%
\author{Meredydd Williams\inst{1} \and
Kelvin K. K. Yao\inst{1} \and
Jason R.C. Nurse\inst{2}}
\authorrunning{Williams, Yao \& Nurse}
%
\institute{University of Oxford, UK\\ 
\email{meredydd.williams@cs.ox.ac.uk}\\
\and
University of Kent, UK\\
\email{j.r.c.nurse@kent.ac.uk}}

\titlerunning{Augmented Reality Tourism Apps through User-Centred Design}
\maketitle              
\begin{abstract}
 Augmented Reality (AR) bridges the gap between the physical and virtual world. Through overlaying graphics on natural environments, users can immerse themselves in a tailored environment. This offers great benefits to mobile tourism, where points of interest (POIs) can be annotated on a smartphone screen. While a variety of apps currently exist, usability issues can discourage users from embracing AR. Interfaces can become cluttered with icons, with POI occlusion posing further challenges. In this paper, we use user-centred design (UCD) to develop an AR tourism app. We solicit requirements through a synthesis of domain analysis, tourist observation and semi-structured interviews. Whereas previous user-centred work has designed mock-ups, we iteratively develop a full Android app. This includes overhead maps and route navigation, in addition to a detailed AR browser. The final product is evaluated by 20 users, who participate in a tourism task in a UK city. Users regard the system as usable and intuitive, and suggest the addition of further customisation. We finish by critically analysing the challenges of a user-centred methodology.

\end{abstract}
\section{Introduction}

Augmented Reality (AR) projects virtual graphics into real-world environments, thus bridging the gap between the physical and virtual spaces. These interfaces can enhance human vision, helping users absorb information in an intuitive manner \cite{Zhou2008}. AR systems offer benefits in a variety of fields, from healthcare to marketing, manufacturing to entertainment. Such tools are particularly useful for navigation, enabling tourists to move away from static paper-based maps \cite{Abowd1997,yung2019new}. Points of interest (POIs) can be highlighted by overlaid icons, helping users negotiate unfamiliar cities. While such systems previously required head-mounted displays, smartphones can now support this functionality \cite{Allison2015}. It is worth noting that smart devices are increasingly researched in mixed-reality environments \cite{park2019new}; this is despite their related privacy issues \cite{williams2019smart,williams2019smartwatch}. With such mobile devices pervading our lives in several cases, tourism apps have grown significantly in popularity \cite{ramos2018new}.

While AR offers advantages to tourists, apps have been criticised for their usability issues. Julier et al. (\cite{Julier2000}) found that displays are often cluttered, while Tokusho and Feiner (\cite{Tokusho2009}) highlighted the limited field of view. Occlusion can occur when overlays collide, causing confusion as icons are obscured. Many apps require the device to be held constantly upright, resulting in an uncomfortable stance. Usability is a goal of user-centred design (UCD), where feedback is sought throughout the development process \cite{Vredenburg2002}. Through soliciting requirements and refining prototypes, the product is often better-suited to users' needs.

Our contribution is the user-centred development of an AR tourism app. We first extract our requirements from a synthesis of domain analysis, interviews and tourist observations. This allows user suggestions to be informed by existing best practice. We then proceed through four rounds of iterative prototyping, refining our overhead map, route planner and AR browser. Rather than developing design frameworks, as has been done in previous work \cite{Olsson2011,han2019virtual}, we implement a full Android application. To empirically evaluate our system, we conduct live trials with 20 participants. These users judge our app to be usable and intuitive, and recommend the addition of further customisation. We finish by critically analysing the challenges of a user-centred methodology. 

The rest of our paper is structured as follows. Section 2 reflects on related work, including usability studies and AR apps. We describe our requirement-gathering process in Section 3, which synthesises domain analysis, user interviews and tourist observations. In Section 4 we outline our development process and the iterative prototypes. Section 5 concerns our user evaluation, study findings and methodological challenges. We conclude in Section 6 and discuss opportunities for future work.

\section{Related Work}

\subsection{User-Centred Design}

Before we discuss our development, we should clarify what is meant by user-centred design (UCD). Vredenburg et al. (\cite{Vredenburg2002}) define it as ``\textit{the active involvement of users for a clear
understanding of user and task requirements, iterative design and evaluation, and a multi-disciplinary approach}''. The process involves participatory design where users do not simply evaluate a product on completion, but deliver feedback throughout development. The general approach is summarised in Roda~\cite{Roda2014}.


\subsection{AR Tourism Literature}

AR apps have been studied through a range of works. Grubert et al. (\cite{Grubert2011}) conducted an online survey to ascertain browser opinions. Usability issues were frequent, with respondents frustrated by the awkward stance for holding their phones. Whereas this survey identified issues, we extract requirements for a real application. 

Olsson and Salo (\cite{Olsson2011}) undertook focus groups and online surveys to study user expectations. Participants expressed that AR should increase the efficiency of everyday tasks. Users also believed discovery should be facilitated through an engaging environment. However, while the researchers only extracted system requirements, we develop a user-centred Android app.

In other work, Yovcheva et al. (\cite{Yovcheva2015}) explored AR interface design. They created a range of annotation mock-ups, before evaluating these through a user study. Participants were found to value names and descriptions, with walking time considered less important. They also evaluated how to highlight POIs most effectively. The researchers discovered that colour-coded overlays were most salient, but that the technology required is computationally expensive. While these studies inform AR design, they did not result in an implemented system.

Tokusho and Feiner (\cite{Tokusho2009}) developed an AR equivalent for Google StreetView. They encountered several usability issues, including cluttered screens and a limited field of vision. Although the researchers implemented a prototype, their requirements were not user-informed. In contrast, we seek to improve usability through a user-centred approach.

Schinke et al. (\cite{Schinke2010}) suggested 3D arrows to assist navigation. While user studies indicated these shapes were beneficial, the test included only four POIs per screen. Since Fr{\"o}hlich et al. (\cite{Frohlich2006}) found that radars can be complex, maps could be a consideration for our design. We now move forward to outline our requirement-gathering process.

\subsection{Mobile Tourism Apps}

With the barriers to global travel decreasing, tourism apps \cite{Abowd1997} have become increasingly popular. Google Maps, for instance, possesses many features beneficial for tourists. Users can search for a variety of attractions, including restaurants and bars, with these POIs (points of interest) annotated on the map. Foursquare offers similar functionality, but can be customised around the user's preferences. By adding tags corresponding to interests, the application will recommend local attractions. TripAdvisor differs from Google Maps in that it is specifically designed for tourists. The app includes user-generated travel guides, location-specific forums and in-depth restaurant filtering.

While these apps are popular, Augmented Reality offers several advantages. POIs are easier to locate when their position is highlighted in the physical world. Furthermore, navigation is enhanced when users know in which direction their destination lies \cite{Kounavis2012}. Unfamiliar attractions might be hard to locate, even once users near the vicinity. Annotations allow the POI to be clearly marked, with the phone's bearing indicated by its magnetometer.

With AR offering advantages over conventional apps, these tools have grown popular. In this work we constrain our focus to Android applications, as they offer greater flexibility in AR design. Wikitude (\cite{Wikitude2017}) is an Austrian technology provider which offers both AR browsers and development kits. Their app interfaces with Google Places, populating a locale with nearby attractions. Despite useful features, the application possesses usability issues. In an attempt to reduce occlusion, annotations are grouped when several overlap. While this reduces on-screen cluttering, the icons may be too small for simple use. With user orientation being key for navigation, the current location should always be displayed. However, Wikitude often places POIs over this symbol, thus causing potential confusion.

ARNav (\cite{ARNav2017}) offers similar functionality, allowing users to search for nearby attractions. Rather than crowding the screen with icons, the app presents a list of POIs to be selected. The system can even identify mountains, with their distance calculated from the GPS position. However, the app does suffer from usability issues, with POI selection being a cumbersome process.

\section{Requirement Gathering}

\subsection{Method}

To create our AR application, the first task was to define our system's requirements. Given that the aim of this research was to develop a highly-usable app, we adopted a strict user-centred process. This stage comprises the `User Research' component of User-Centred Design, as shown in Roda~\cite{Roda2014}. In extracting our product requirements, we used a synthesis of approaches. Firstly, we built on prior work to establish the fundamentals of an AR system. As non-technical users might not be familiar with AR browsers, they cannot be expected to design the application unassisted. With domain analysis extracting existing best practice, participants can select requirements with greater feasibility. 

Secondly, we conducted a participant observation with real tourists. We selected this study as it supported data extraction from a real-life setting \cite{Spradley2016}. Through surveying behaviour, we identified how people navigate a location and what information they desire. This ensured requirements reflected the needs of ordinary tourists. Informal conversations informed a set of interview questions, ensuring we collected the most-relevant information. 

Finally, we undertook these interviews with 14 participants. Through semi-structured discussions of AR design, we considered important features and the decision-making process. This qualitative data enriched the observation notes and existing best practice, supporting the construction of our requirement list. All the requirement-gathering tasks were ethically approved by our IRB board. These processes are discussed in the rest of this section.

\subsection{Domain Analyses}

We first investigated AR tourism apps to identify their common components. A product which is not informed by current expertise might either `reinvent the wheel' or lack feasibility. By supplementing best practice with user requirements, we can develop applications which are both feasible and usable.

To begin our domain analysis \cite{Zowghi2005}, we examined applications in the Google Play Store. We selected this portal as we construct an Android app, with the search phrase of `\textit{augmented reality tourism}' used. We retrieved 192 applications, 73 of which concerned AR with 45 of those in the English language. Considering those 22 with user ratings of 4/5 and above, we identified the features which individuals appreciated. Common functionality included contextual information, usable controls and a clean interface. Most browsers included a built-in map to display nearby POIs. Attractions should also be enhanced with helpful data concerning type, distance and user reviews. This can vary based on the POI, such as lunchtime menus for restaurants.

We then surveyed user studies and the significance of their findings. One criticism of user-centred design is that the product is excessively customised to idiosyncratic opinions \cite{Abras2004}. By supplementing our requirements with academic expertise, our browser should be suitable for a range of users. Whereas apps should not arbitrarily omit POIs from the display, visual cluttering is a frequent issue \cite{Olsson2011}. Yovcheva (\cite{Yovcheva2015}) found the most-recognisable annotations used colour-coded highlighting. However, the resource constraints of smartphones can make computer vision infeasible. 

In addition to conducting user studies, Yovcheva (\cite{Yovcheva2015}) provided an AR overlay framework. It consisted of eight usability points, such as prioritising details for nearby attractions. We also surveyed generic usability guidelines, such as those developed by Schneiderman (\cite{Schneiderman2009}) and Nielsen (\cite{Nielsen1994}). We factored these points into our final requirements. 

\subsection{Tourist Observation}
\label{sec:threec}

After identifying best practice from previous work, we solicited opinions from ordinary tourists. Whereas design frameworks can form a foundation, requirements must be informed by ordinary users. We selected interviews as our main means for extracting comments. However, we required contextualised data to construct our interview questions. Through observing the activities of ordinary tourists, we were able to define our queries. As our interviews also included a user scenario, our observation also informed the design of this task.

Our study was conducted with six tourists in Oxford, UK. These participants were recruited in situ from members of the public. We observed their navigational behaviour and the resources they used during their visit. Whereas some tourists used smartphones for research, many relied on maps and guidebooks. After observation, we informally solicited the participants' experiences. Topics included navigational difficulties and the factors that influence their decisions.

All six of our participants had planned their own trips to popular attractions. Through discussions of surrounding landmarks, we found the tourists generally knew little about the POIs. They were more encouraged by photos and user reviews than by the history of iconic buildings. Several expressed that while they wished to learn more about the sites, they had no access to this information. We also found many participants needed a coffee shop's WiFi to learn about the local area. This implies that apps with offline content could be of benefit.

Our discussions highlighted the importance of pre-trip planning. We also found a variety of paper-based and digital methods were used to obtain on-site details. Since route planning, features (e.g., top attractions, etc.) and decision-making (e.g., what should I visit?) seemed most important, we constructed our interview questions around these topics. Several suggestions had merit, such as offline content, and these were added to our requirements.

\subsection{User Interviews}
\label{sec:threed}

Informed by the topics of Route Planning, Features and Decision Making, we developed our interview questions. These included queries about user's preferred applications and favourite features. The interviews were contextualised around a user scenario, grounding discussions in a tourism environment. This helped participants consider a real trip, rather than answering queries in a disengaged manner. The scenario was informed by our observations, building on the common activities mentioned by tourists. The scenario is as follows:

\begin{quoting}
`\textit{Richard is planning to tour X. It is the first time that he will be visiting the city and he knows nothing about it. When he arrives at X, he will explore around the city centre. He will use his smartphone to locate places to eat and attractions to visit. He is also fascinated by the historical buildings and wishes to learn more information. Finally, he must return to the train station before his train departs}'.
\end{quoting}

We selected a semi-structured approach to further explore participant responses. All 8 questions were open-ended in nature, encouraging respondents to explain their answers in detail. This qualitative data helped construct an extensive list of requirements.

We conducted interviews with 14 participants, with the duration ranging from 30 to 45 minutes. These users were recruited from students at a local university. Findings are grouped based on our three aforementioned topics: Route Planning, Features and Decision Making. 

Participants were asked whether they plan prior to their trip and, if so, what techniques do they use. 13/14 respondents organised their tour, with all of these reporting that they made use of the Internet. Most individuals used Google Maps to mark attractions, before constructing an itinerary on paper. Distance to travel was the most important priority, with 8/14 stating they optimise their routes. Our planning discussions suggested that an AR tourism app would be appreciated. By consolidating reviews, routes and POI details, tourists could navigate locations in an informed manner.

Participants next ranked six features in order of importance. These comprised navigation, events, route creation, local restaurants, suggested routes and top attractions. Individuals were also invited to suggest other features they would find valuable. The highest-rated functions were top attractions and local restaurants, with events deemed least important. Participants were also asked which details they require about a POI. Most requested data such as price, photos and opening times. Suggested features included nearby toilets and optimised routes, with these points added to our requirements.

To assess decision-making, participants were tasked to select a local attraction. They disclosed the factors which influenced their choice, listed in order of importance. We found distance and user ratings were most influential, though price was mentioned on several occasions. Through our semi-structured interviews we identified a number of requirements. Using our user-centred approach, we believe these will contribute to a usable and functional AR app.

\subsection{User-Centred Requirements}

Our user-centred requirements have been extracted through three processes: domain analysis, tourist observations and semi-structured interviews. The final list of requirements was divided into five themes: POI Information, Route Planning, Interface Design, AR Design and Miscellaneous. We summarise the requirement topics below:

POI Information concerned the data presented for each attraction. Based on the most popular factors from both prior work and our user studies, we prioritised details for our POIs. Route Planning was highly valued, whether conducted before the trip or on-site. The application should allow users to select their POIs and follow an optimised path. Interface Design is critically important to ensure the application is usable. Our principles, informed by both prior studies and user comments, include clarity, consistency and coherence. By minimising the cognitive load on tourists, we can support their requests in a usable manner. 

AR Design contains several challenges, including visual cluttering and off-screen content. As users are usually more interested in nearby POIs, annotations can be grouped and sorted by proximity. With design decisions requiring user feedback, we trialled different versions through our prototypes. Miscellaneous requirements came from user suggestions, such as offline content and nearby toilets. With these features proposed by real tourists, we expect the functionality will benefit our target audience.

\section{Iterative Development}

\subsection{Method}

After extracting our user-centred requirements, we began a process of iterative development. Within this methodology (as outlined in Roda~\cite{Roda2014}), we refine prototypes through frequent user testing. After each round we feed participant comments back into development, leading to an app which should match user expectations. 

Through our user-centred approach, we solicited the opinions of 10 participants. These users were again recruited from students in a local university. Feedback was audio-recorded, with the process ethically approved by our University's IRB. Our development consisted of five stages: Initial Mock-ups, Map Prototypes, Route Planner Prototypes, AR Browser Prototypes and the Final Design. We now briefly outline each phase in turn.

In Initial Mock-ups, we created low-fidelity designs for the user interface. These mock-ups included the main menu screens and the AR browser annotations. With these elements comprising key functionality, we felt it appropriate they were designed first. After soliciting feedback, we moved forward to build Map Prototypes. The overhead map was a crucial component, with AR building on this functionality. Before AR features were implemented, it was essential we developed the underlying maps. 

The Route Planner, suggested by several tourists, is a key component of both Google Maps and TripAdvisor. In wishing to provide an app of similar functionality, we refined our design through iterative development. We developed AR Browser Prototypes last as the navigation tools required a strong foundation. Following iterative development and repeated consultation, we delivered our Final Product. We now proceed through each stage in detail, describing the prototype, the feedback and our user-centred approach.

\subsection{Stage 1: Initial Mock-ups}
\label{sec:fourb}

We began by creating a series of low-fidelity mock-ups, as found in the `Design' component of UCD (see Roda~\cite{Roda2014}). These were designed to give our users an overview of the future application. Rather than building a full system, which might be costly to adjust, mock-ups allow cheap and rapid refinement. We first sketched the main menu screen, which users will navigate to access the app's functionality. 

\begin{figure}[h!]
    \includegraphics[width=0.8\textwidth]{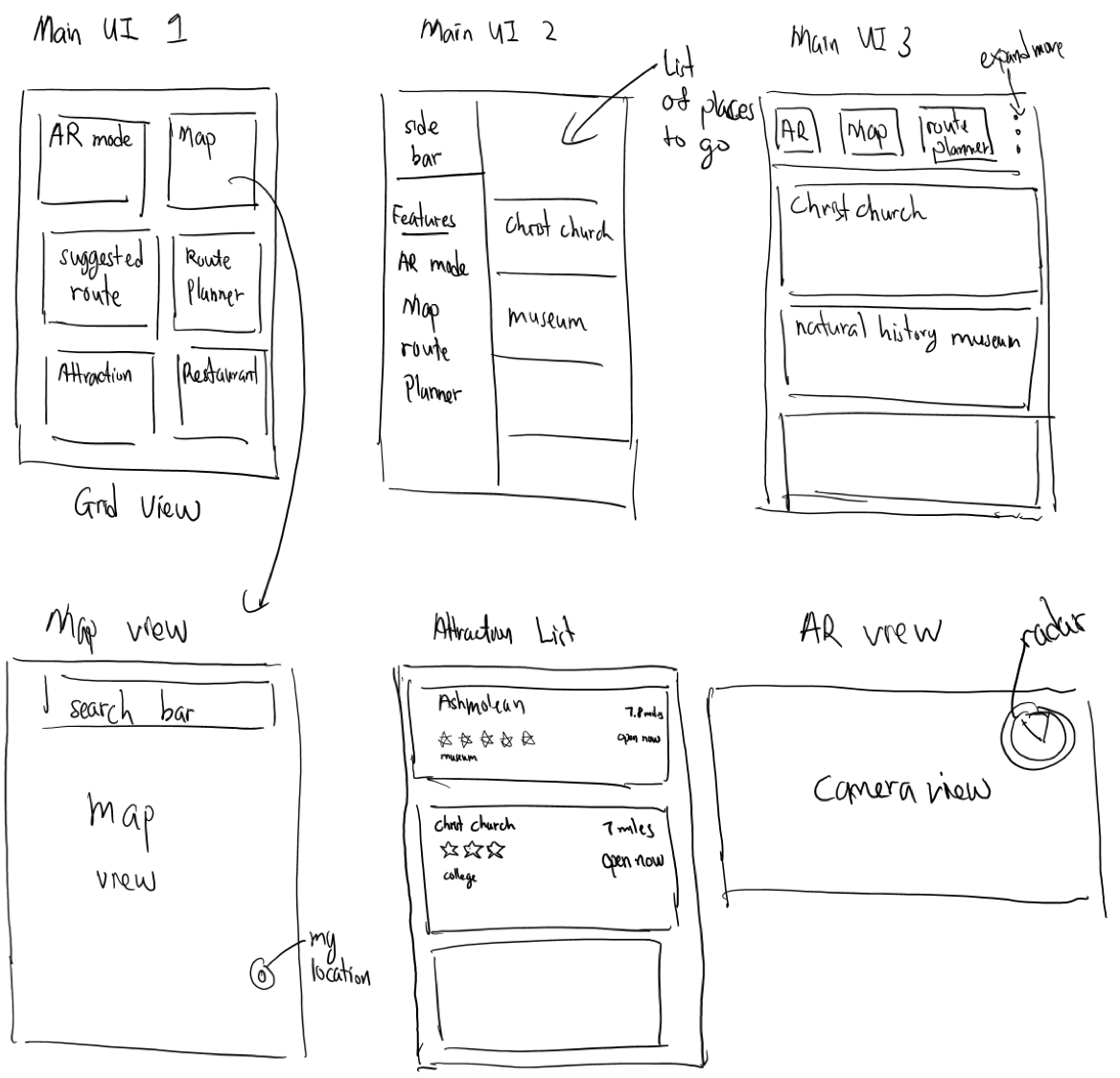}
    \centering
    \caption{Mock-up interfaces: Main UI 1 (top-left), Main UI 2 (top-centre), Main UI 3 (top-right), Map View (bottom-left), POI List (bottom-centre) and AR View (bottom-right)}
    \label{fig:mockup}
\end{figure}

The mock-ups were designed around four principles drawn from our requirements. We wanted to build an app which was intuitive and easily understandable to our users. We also required the interface to allow quick navigation between different features. For our app to be usable, this navigation must be simple and consistent. Finally, to minimise the cognitive load on users, the interface should be clean and coherent. The low-fidelity mock-ups are presented in Figure \ref{fig:mockup}.

Our participants analysed the interface sketches and gave qualitative feedback. They decided that `Main UI 1' was most preferable, as it lists all the features on a single screen. They also appreciated `Main UI 3', where popular attractions are displayed on the top bar. A user commented: ``\textit{The UI is more attractive with images of the attractions and I can find all the attractions. There is also a top bar showing other functions of the app}''.

Whereas `Main UI 3' advertises POIs, it does obscure other pieces of functionality. In contrast, `Main UI 1' showcases the wide range of available functions. With both interfaces appreciated and the menu required for navigation, `Main UI 1' was selected for the launch screen.

\begin{figure}[h!]
    \includegraphics[width=0.9\textwidth]{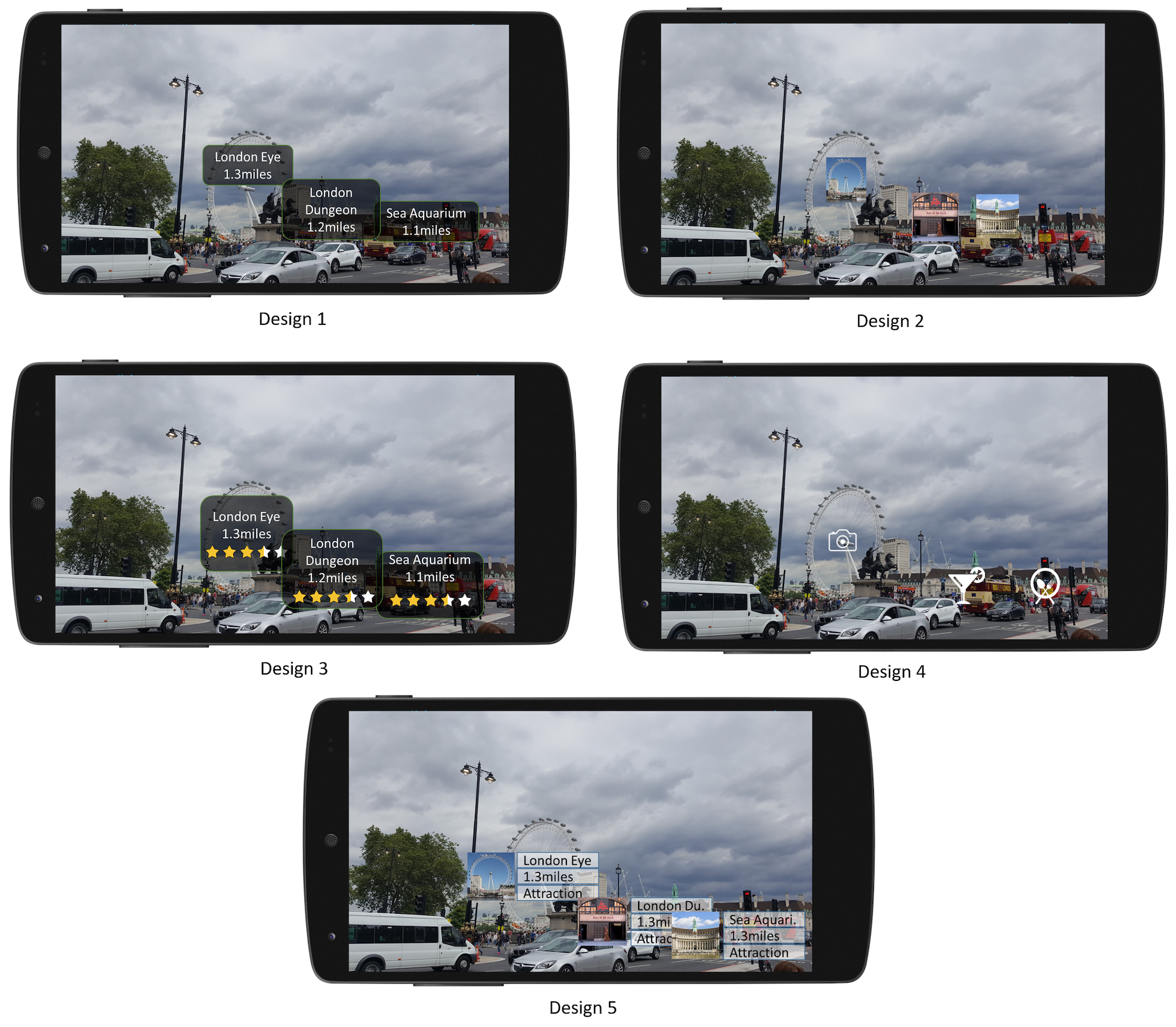}
    \centering
    \caption{Mock-up annotations: Design 1 (top-left), Design 2 (top-right), Design 3 (centre-left), Design 4 (centre-right) and Design 5 (bottom)}
    \label{fig:interfaces}
\end{figure}

We next developed 5 mock designs for the AR annotations. Diagrams were implemented using stock images and overlaid icons. These designs varied around four factors: size, recognisability, details and decision-making information. Whereas one annotation included a small image, another used name and distance. Our users were presented with these candidates and asked which they preferred. The alternatives are presented in Figure \ref{fig:interfaces}.

Most participants found Design 3 most useful, which contained POI name, distance and user rating. They also appreciated Design 4 as the icon indicated the type of attraction. This was important as a POI's name might be irrelevant if the category is not known. Many participants commented that images should not be included, as they occupied too much space on the user interface. One mentioned: ``\textit{If there are many attractions in one area, there will be a mess if many pictures are displayed on the screen}''.

8/10 participants agreed that distance helps them locate the attractions. They also commented that user ratings can help them filter out undesirable locations. Based on this feedback, we selected four characteristics for our POI annotations: name, type, distance and rating. By subscribing to this user-centred approach, we believe the implemented interface will be more usable.

\subsection{Stage 2: Map Prototypes}

\begin{figure}[h!]
    \includegraphics[width=0.9\textwidth]{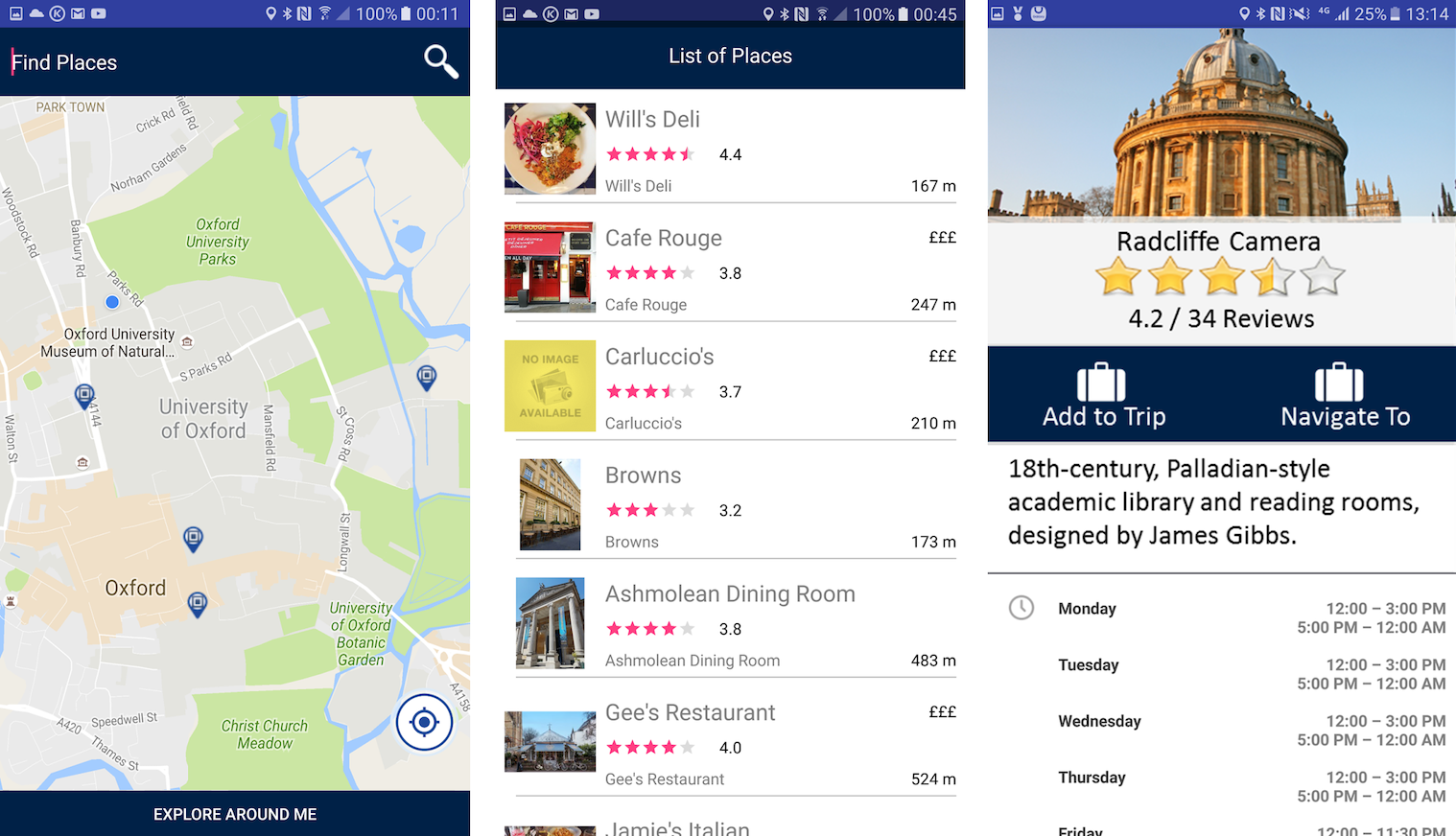}
    \centering
    \caption{Map prototype: Overhead View (left), POI List (centre) and POI Details (right)}
    \label{fig:map}
\end{figure}

Prior research suggests maps are the best technique for spatial visualisations \cite{Kraak2011}. AR acts as an enhancement to this functionality, placing helpful annotations on the physical world. Without an overhead projection at least in the internal system, the annotations lose their spatial relevance. Therefore, before we developed AR features, we prototyped the underlying maps. This comprises the `Build' component of User-Centred Design, as shown in Roda~\cite{Roda2014}. We developed a prototype Android application, incorporating data from Google and Wikipedia APIs. The initial prototype is presented in Figure \ref{fig:map}.

Feedback suggested our interface was usable, with some caveats. One user searched for an attraction but received no results due to a typographic error. We therefore inferred that an autocomplete function might assist individuals. Users complained about interface navigation, with moving between POIs proving cumbersome. Individuals had to press the back button and then select another attraction, with this causing some frustration. This was clearly an area which required additional work.

Based on user feedback, we implemented several revisions. By developing an autocomplete feature, individuals were assisted in their search for attractions. `BottomSheets' were implemented to improve navigation, with the interface revealed when the user swiped upwards. This helped individuals browse details without moving to another screen. Based on our user-centred approach, we believe these revisions improved the app's usability.

\subsection{Stage 3: Route Planner Prototypes}

\begin{figure}[h!]
    \includegraphics[width=0.9\textwidth]{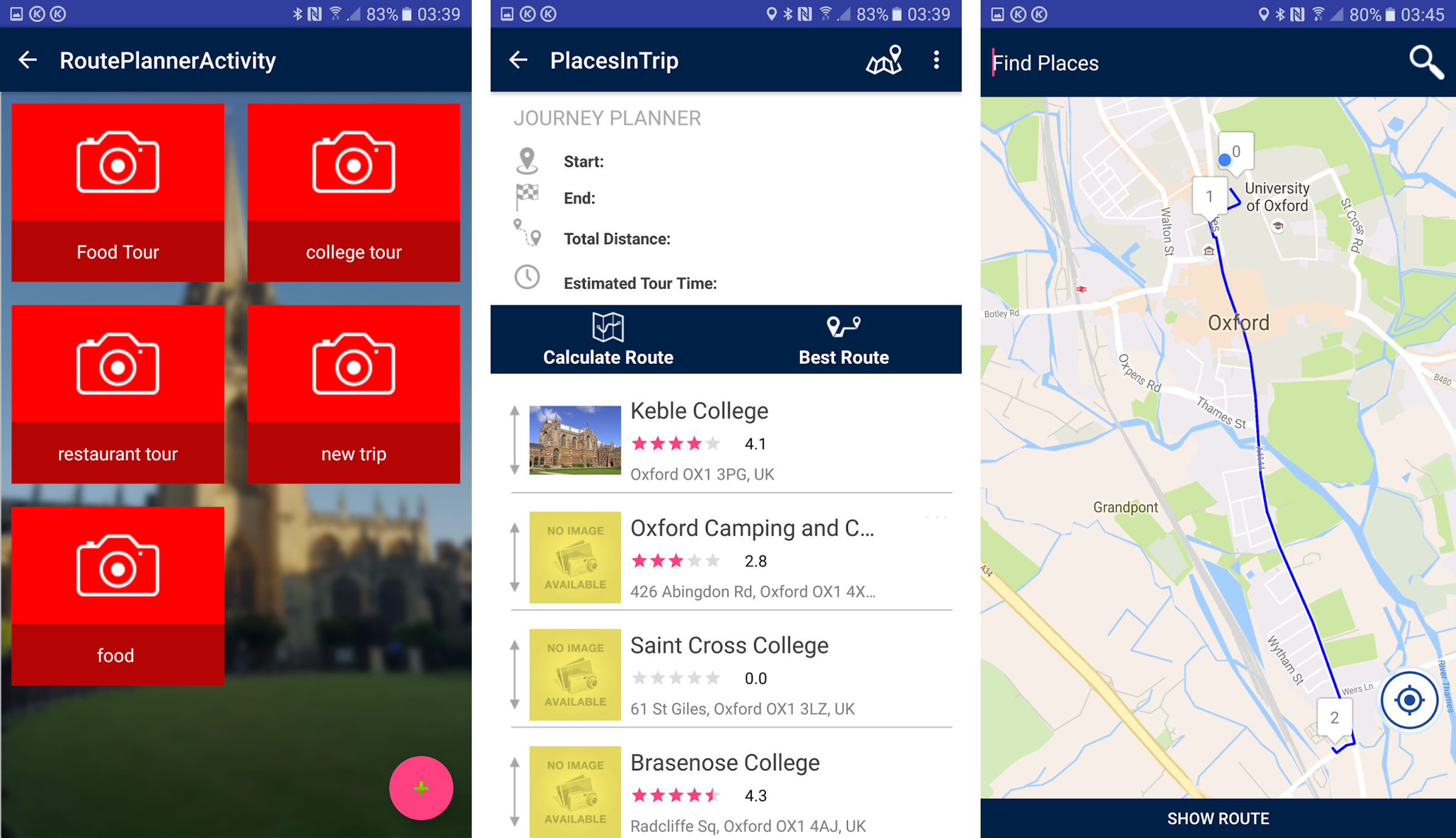}
    \centering
    \caption{Route planner prototype: Trip Selection (left), Trip Creation (centre) and Route Navigation (right)}
    \label{fig:route}
\end{figure}

In our tourist observation, presented in Section \ref{sec:threec}, we found that route planners were greatly appreciated. These tools enhance map functionality, enabling users to navigate in an efficient manner. With our app designed for tourism purposes, we believed route planning was required at an early stage. Conforming to our user-centred approach, we developed and evaluated an initial prototype. The design was informed by Schneiderman's `Golden Rules' (\cite{Schneiderman2009}), including consistency and informative feedback. The app is shown in Figure \ref{fig:route}.

Individuals create a new trip by selecting a collection of POIs. The user then uses the route planner to navigate themselves between attractions when they reach the city. We designed icons for each POI type, in a similar manner to prior work \cite{Yovcheva2015}. For example, restaurants were represented by a knife and fork, while bars were denoted by a cocktail glass. To assess the usability of the system, we requested feedback from our participant group.

Several users noted that once POIs were added to a trip, there was no functionality to remove them. In this case, individuals needed to create a new trip and add the attractions individually. As tourists might arrive in cities from different directions, users asked for the POI sequence to be adjustable. Similarly, the start and end locations were initially defined based on the attractions. Users requested the ability to set a defined end point, such as a hotel or train station.

We refined our prototype based on the participant feedback. We first implemented POI checkboxes, allowing them to be easily added and removed from trips. Start and end locations also became adjustable, with navigation routes recalculated automatically. Routing algorithms were further enhanced, with paths considering opening times and duration of visit. Therefore, routes now more-accurately reflected the practicalities of undertaking a trip. With the map foundations now finalised, we move forward to implement AR functionality.

\subsection{Stage 4: AR Browser Prototypes}

In previous stages we prototyped interfaces, maps and route planners. These are the elements found in most tourism apps, including TripAdvisor and Google Maps. While a purely AR application could assist localised navigation, tourists might have difficulty finding distant attractions. By building on our user-refined features, we sought to provide a usable app.

\begin{figure}[h!]
    \includegraphics[width=0.9\textwidth]{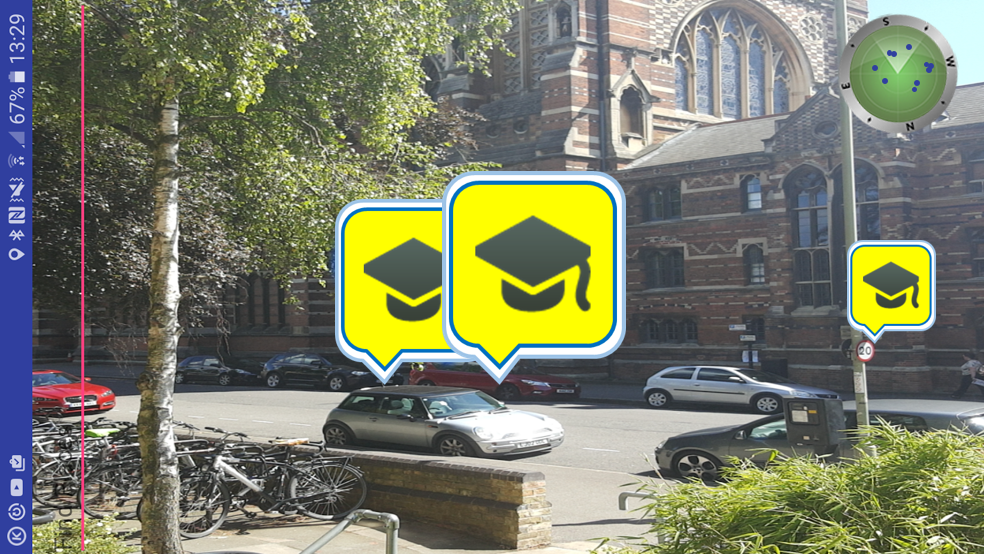}
    \centering
    \caption{Initial AR browser prototype}
    \label{fig:ar1}
\end{figure}

We began by prototyping a basic browser, which allowed users to search for local POIs. Attractions were represented by a simple icon, with a search bar and radar also included. While icon annotations were less popular in our mock-ups (Section \ref{sec:fourb}), we trialled this approach for quick feedback. Our initial AR prototype is shown in Figure \ref{fig:ar1}.

Participants commented that annotation size should represent proximity, with larger icons denoting nearby attractions. It was observed that individuals usually held their phones in a portrait orientation, rotating to landscape to use the AR browser. They would then return to portrait mode to read the POI details, causing frustration. Users also suggested that we add category shortcuts to the search bar. For example, rather than typing ``bars'' in a cumbersome fashion, they would prefer to select a cocktail icon.

Reacting to user feedback, we implemented AR in portrait mode. If individuals rotated their device to landscape, the display would also adjust accordingly. This would help tourists operate the app with one hand, leaving the other free for maps or luggage. We also implemented search bar shortcuts, simplifying the selection of bars and restaurants. To user test our annotations, we replaced our icons with three alternatives. `Simple text' contained name, type and distance; `Detailed text' also possessed description and rating; while `Image' consisted of a POI photo.

Most participants were pleased with the browser updates. The majority of users preferred the `Detailed text' annotations, as it provided a range of information. With this approach being quite verbose, some suggested that icons could be used when the screen becomes cluttered. Others commented that the annotations were too small and this impeded the recognition of POIs. One participant mentioned: ``\textit{Detailed-text design is good enough, although having a photo is good if the place is not visible, but I think it is too small on the screen}''.

To accommodate this feedback, we allowed users to adjust the overlay size. With occlusion and crowding causing confusion, we also developed a grouping mechanism. When AR overlays begin to overlap, they are collected into a scrollable list. Users can navigate this list by clicking the annotation, with grouped attractions ranked in proximity order.

Through prior work \cite{Julier2000}, we found that AR usability is frequently criticised. The awkward stance required to use these apps might discourage users. With AR serving little purpose when the phone faces downwards, we configured the interface to switch modes. This enables the user to move seamlessly between the browser and the overhead map through a simple gesture. After presenting these design updates to our user group, all 10 were pleased with the app's functionality.

\subsection{Stage 5: Final Design}

\begin{figure}[h!]
    \includegraphics[width=0.9\textwidth]{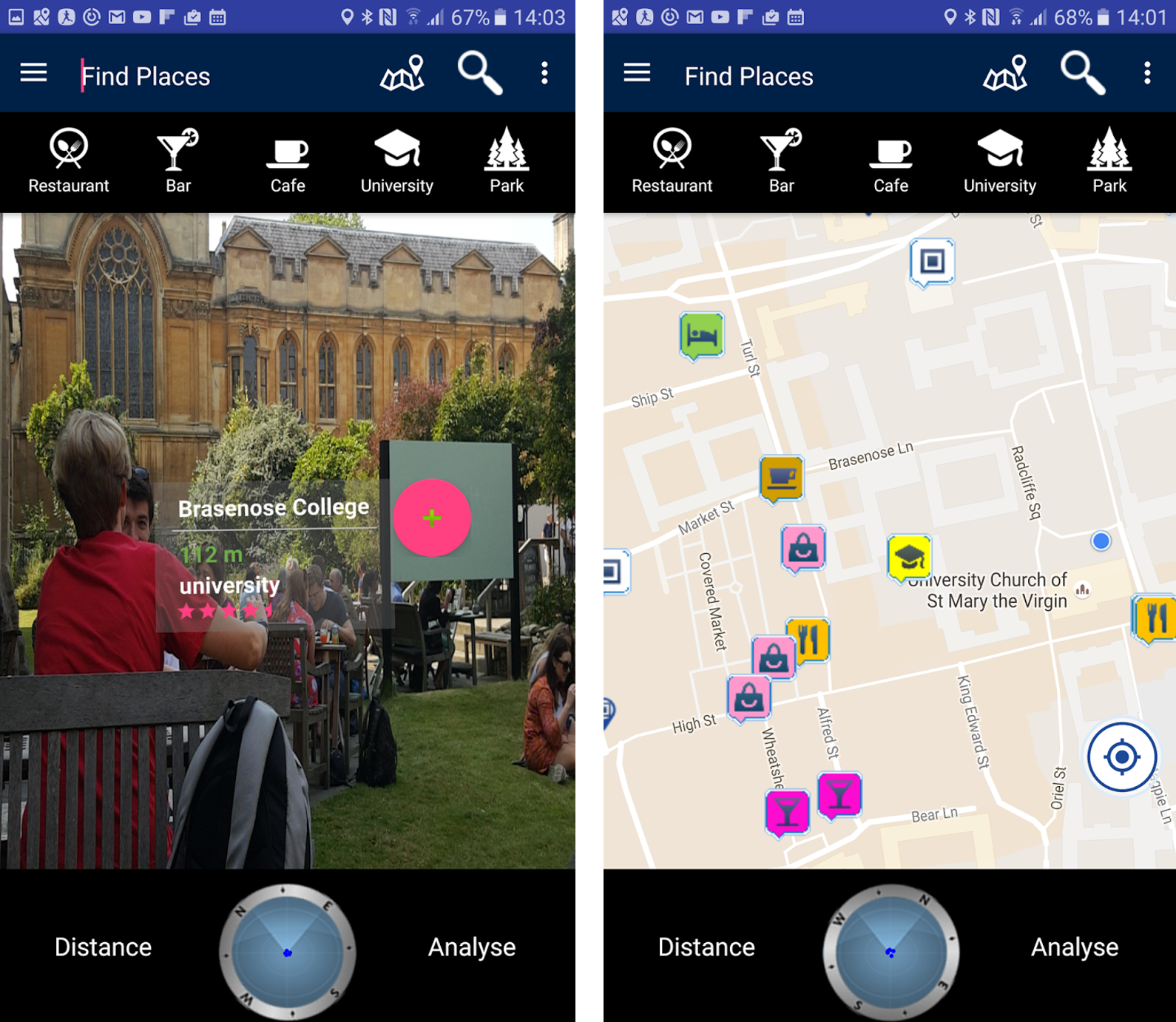}
    \centering
    \caption{Final AR tourism application: AR Browser (left) and Overhead Map (right)}
    \label{fig:final}
\end{figure}

After four rounds of prototyping, evaluation and user-centred refinement, we constructed our final application. Before completion, we added one additional feature: offline content. This was not developed through our user testing, as our study participants had Internet connections. However, we recognised that many tourists will not have access, as was found during our observations (Section \ref{sec:threec}). After route trips and POIs have been selected, they remain navigable without Internet access. The final app is displayed in Figure \ref{fig:final}, which showcases the switching between the AR browser and the map.

This application has been tested and refined through an iterative process. However, piecemeal feedback does not provide the rigour required for a comprehensive evaluation. Thus far, individual features have been tested in isolation. To ascertain whether the app could benefit ordinary tourists, we move on to evaluate the system in its entirety. This comprises the `Usability Testing' element of User-Centred Design (see Roda~\cite{Roda2014}), where we test whether the solution meets the needs of the users.

\section{Evaluation and Discussion}

\subsection{User Evaluation Method}

We recruited 10 participants through our user-centred development. While we considered using the same sample, we wished to explore the wider applicability of their opinions. A distinct user group would have sufficed, but we also desired validation of our requirements. To both validate our design and evaluate at a larger scale, we recruited 10 additional students. By using a test sample of 20 users, we evaluated the app through a range of viewpoints.

To test the system in a real-world environment, we developed a tourism exercise. In this task, users were observed while visiting POIs around Oxford, UK. When a participant reached one of our defined stops, they were asked to use the app to explore the area. The open-ended scenario was as follows:

\begin{quoting}
``\textit{Jane is going to visit X. She wants to plan her route but only has a single day to explore the city. Jane will use the app to undertake route planning. When she does arrive at X, she will use the app to navigate the area. She will also use the AR browser to explore attractions and receive more information on the POIs}''.
\end{quoting}

Participants were first given a brief demonstration of the app. An information sheet was also provided, in addition to the written scenario. Users had the opportunity to trial the app so they could ask any pertinent questions. After participants finished the navigational tasks, the study was completed by an exit questionnaire. This contained 9 questions, concerning functionality, usability and AR design. Participants were given an hour to perform their tasks. Since the route was estimated to take 30 minutes, we believe this was ample time.

\subsection{Study Findings}

Four main themes emerged from our findings: User Experience, AR Browser, Route Planning and POI Annotations. Each of these topics will now be discussed in turn.

In terms of User Experience, participants remarked that the application was simple and intuitive. They commended the consistent use of colour and praised the coherence of the interface. One user particularly enjoyed the AR browser, asking for the app to start on this screen. Since the system's Unique Selling Point was this mode, the alteration could be considered in future updates. General feedback was complimentary: ``\textit{The UI is more or less self explanatory. Overall experience is very good}''.

The AR Browser also received praise, both for its usability and functionality. Grouping overlays was considered successful in reducing on-screen cluttering. Several remarked that they would be more likely to explore their area if they owned such an app. However, one individual had smartphone reservations, claiming that AR could not be adequately supported: ``\textit{I am not sold on AR in mobile context, the user experience is good because of the context approach which is not directly tied with AR}''.

Several users commented that bystanders avoided their field of vision, believing that the participant was taking a photo. Being cognisant of this fact, we ensured all the stops were historical attractions. Therefore, a person holding their phone in this manner should feel less awkward. Despite these generic AR issues, users praised the gesture navigation between browser and map. One stated: ``\textit{I like how you simply flip the phone to use AR function. It is good to have both AR and map view let users decide which they prefer}''.

Rather than concerning design, the most frequent complaint came from phone hardware. Several users encountered issues with magnetic interference, which reduced the accuracy of their magnetometers. This caused the AR annotations to flicker, impacting browser usability. While the overlays were appreciated, adjusting their size was found challenging. One user suggested a pinch gesture could be used, and that this should be added in updates.

The Route Planner received consistent praise, with the functionality commended. Participants appreciated the offline content, which allowed navigation without an Internet connection. The routing algorithm was also complimented, as it factored opening times into its calculations. Participants did offer suggestions for upcoming releases, including pop-up notifications and automatic end-points. As one noted: ``\textit{It would be better if it can pop up notification to show information about the route, the next place to go, the distance to next attraction etc}''.

Finally, the POI Annotations were also praised. Most users preferred the `Text and review' overlay, which included name, rating and attraction type. Although several participants appreciated images, the majority believed the details were sufficient. Icons were found to present a high-level overview of nearby attractions. However, since a downward gesture opened the map, many preferred this overhead view.

Our participants appear to regard the application as both functional and usable. We move on to discuss general themes from our comprehensive evaluation.

\subsection{Discussion}

During the final task, it was observed that the style in which tourists navigate can be grouped into three categories. Some individuals planned their trip in detail and visited the POIs in the most efficient manner. These users are goal-driven and less interested in exploring unscheduled attractions. They benefited most from the route planner, which plotted a shortest path between the locations. 

Other individuals registered the POIs but navigated the city in a flexible manner. They structured their trip around certain attractions but preferred the scenic route to the shortest path. These users demand an app which not only provides the direction, but takes them via other POIs. Alternatively, some individuals were exploratory, locating attractions in situ. These participants used the app while walking, and therefore required accurate annotations. A successful AR tourism app must cater for the needs of all three groups.

Users appreciated our technique to reduce occlusion: ranking intersecting overlays by their proximity. However, annotations still often cluttered the screen. When POIs are not nearby, icons could be used to advertise the range of attractions. Although our participants claimed that images aid recognition, most believed them to be superfluous. Since distance information was embedded in the browser, users also used these details to locate the POIs. As bearing readings are prone to magnetic interference, this data could prove helpful to disorientated tourists.

The transition gesture proved very popular, with it enabling quick movement between the map and browser. This suggests that users appreciate using both tools to locate attractions. While AR technology is exciting, it should only be used where it adds value. If users visualise distant POIs better on a map, usability is impaired if this choice is obstructed. With screen real estate limited on smartphones, AR designers should further explore the role of gestures.

Participants considered the route planner to be their favourite feature. This could be due to apps generally possessing AR or routing algorithms, but not both. Although ARNav (\cite{ARNav2017}) has these features, its inconsistent interface might deter tourists. In contrast, our usability is refined through an iterative user-centred process. Despite its popularity, the route planner was not used to its full potential. Advanced settings, such as defining duration at each POI, were not configured by any participant. This might be because such features were hidden in submenus to reduce visual cluttering. Users had difficulty using the planner before the app demonstration, suggesting its usability could be enhanced. While we subscribe to Android design principles \cite{Google2017}, we could ensure our interface reflects familiar apps such as Google Maps.

\subsection{User-Centred Challenges}

While we believe user-centred design offers many advantages, we would like to highlight our encountered challenges. Users often requested the perfect system: one that is functional, performant, usable and attractive. Although it was opaque to our participants, trade-offs were often required at the design level. Requests were also made which were contradictory, particularly if they originate from different users. While one participant wanted larger annotations due to poor eyesight, another preferred smaller symbols. In these situations we went with the majority view, unless it would have crippled usability.

Design is highly subjective and different individuals have different preferences. As non-technical users might know little about UI guidelines, there can be a temptation to design by committee. Although most of our participants appreciated the textual annotations, some preferred images. We were required to make an executive decision, as a composite approach would have cluttered the screen. Developing an app without best practice could result in a Frankenstein-esque hybrid of subjective suggestions.

The composition of one's sample should also be considered. Even if demographics correspond to those of the target audience, the opinion of one group might differ to that of another. This was our rationale for inviting 10 additional users for our final evaluation. Aside from design challenges, there were practical issues in frequent consultation. Each round of development was significantly delayed by soliciting feedback, as has been expressed in other works \cite{Abras2004}. While our prototyping stages each took a week to complete, the software development comprised less than half of this time. Participants might also lose interest and withdraw from the process. We encountered initial challenges in recruiting users but were fortunate they did not require replacement.

Despite these challenges, we do endorse user-centred design. Through eliciting feedback throughout the process, we produced an app fulfilling our participant's requirements. With the system informed by tourists and evaluated through a live scenario, we believe it would be appreciated by the broader public.

\section{Conclusion and Future Work}

In our contribution, we developed an AR tourism app through user-centred design (UCD). Rather than defining our own specification, the requirements were informed by ordinary users. This was achieved through a synthesis of domain analysis, tourist observations and semi-structured interviews. In this manner, best practice supplemented the suggestions of our users. We proceeded to construct our novel app through an iterative process of prototyping. Each component was tested in succession, with participant opinions fed back into the design. After completing the app, we evaluated it through a live scenario with 20 participants. Through performing real tourist exercises, our users found the application to be both usable and useful. We finally reflected on the challenges of our user-centred methodology. 

Despite our novel contributions, we accept several limitations to our work. Firstly, while our sample is not insignificant, we would have benefited from more participants. Future work will extend this approach with a larger group of tourists. Secondly, whereas our participants valued the app, their opinions were not made relative to other systems. We would like to compare our tool to popular alternatives, conducting qualitative analyses of usability and performance. 

After considering our user feedback, we developed several suggestions for further work. Our participants praised the gesture which transitioned from browser to the map. Future studies could explore the role of gestures in AR and whether they can simplify cluttered interfaces. With the overhead map preferred for distant POIs, AR should attempt to enhance the experience. This could achieved through 3D isometric projection, with the view updated based on smartphone sensors. Another feature that may be worth experimenting with is improving the situational awareness presented in the tool. There has been research in this area (e.g., \cite{kingston2018using,javornik2019experimental}) which may be incorporated, to provide the user with a more immersive experience. 

%
%
\bibliographystyle{splncs04}
\bibliography{paper}

\begin{thebibliography}{10}
\providecommand{\url}[1]{\texttt{#1}}
\providecommand{\urlprefix}{URL }
\providecommand{\doi}[1]{https://doi.org/#1}

\bibitem{ARNav2017}
{ARNav} (2017), \url{http://arnav.eu/}

\bibitem{Wikitude2017}
{Wikitude} (2017), \url{https://www.wikitude.com/}

\bibitem{Abowd1997}
Abowd, G.D., Atkeson, C.G., Hong, J., Long, S., Kooper, R., Pinkerton, M.:
  {Cyberguide: A mobile context-aware tour guide}. Wireless Networks
  \textbf{3}(5),  421--433 (1997)

\bibitem{Abras2004}
Abras, C., Maloney-Krichmar, D., Preece, J.: {User-centered design}. In:
  Encyclopedia of Human-Computer Interaction, pp. 445--456. Sage Publications
  (2004)

\bibitem{Allison2015}
Allison, P.R.: {Augmented reality business applications start to get real}
  (2015), \url{http://www.computerweekly.com/feature/
  Augmented-reality-business-applications-start-to-get-real}

\bibitem{Frohlich2006}
Fr{\"{o}}hlich, P., Simon, R., Baillie, L., Anegg, H.: {Comparing conceptual
  designs for mobile access to geo-spatial information}. In: Proceedings of the
  8th Conference on Human-Computer Interaction with Mobile Devices and
  Services. pp. 109--112 (2006)

\bibitem{Google2017}
Google: {Android design principles} (2017),
  \url{https://developer.android.com/design/get-started/principles.html}

\bibitem{Grubert2011}
Grubert, J., Langlotz, T., Grasset, R.: {Augmented reality browser survey}.
  Tech. rep., University of Technology Graz (2011)

\bibitem{han2019virtual}
Han, D.I.D., Weber, J., Bastiaansen, M., Mitas, O., Lub, X.: Virtual and
  augmented reality technologies to enhance the visitor experience in cultural
  tourism. In: Augmented Reality and Virtual Reality, pp. 113--128. Springer
  (2019)

\bibitem{javornik2019experimental}
Javornik, A., Kostopoulou, E., Rogers, Y., Fatah~gen Schieck, A.,
  Koutsolampros, P., Maria~Moutinho, A., Julier, S.: An experimental study on
  the role of augmented reality content type in an outdoor site exploration.
  Behaviour \& Information Technology  \textbf{38}(1),  9--27 (2019)

\bibitem{Julier2000}
Julier, S., Lanzagorta, M., Baillot, Y., Rosenblum, L., Feiner, S.,
  H{\"{o}}llerer, T., Sestito, S.: {Information filtering for mobile augmented
  reality}. In: Proceedings of the 2000 International Symposium on Augmented
  Reality. pp. 3--11 (2000)

\bibitem{kingston2018using}
Kingston, C., Nurse, J.R.C., Agrafiotis, I., Milich, A.B.: Using semantic
  clustering to support situation awareness on twitter: the case of world
  views. Human-centric Computing and Information Sciences  \textbf{8}(1), ~22
  (2018)

\bibitem{Kounavis2012}
Kounavis, C.D., Kasimati, A.E., Zamani, E.D.: {Enhancing the tourism experience
  through mobile augmented reality: Challenges and prospects}. International
  Journal of Engineering Business Management  \textbf{4},  10--16 (2012)

\bibitem{Kraak2011}
Kraak, M.J., Ormeling, F.: {Cartography: Visualization of Spatial Data}.
  Guildford Press (2011)

\bibitem{Nielsen1994}
Nielsen, J.: {Usability engineering}. Elsevier (1994)

\bibitem{Olsson2011}
Olsson, T., Salo, M.: {Online user survey on current mobile augmented reality
  applications}. In: Proceedings of the 10th IEEE International Symposium on
  Mixed and Augmented Reality. pp. 75--84 (2011)

\bibitem{park2019new}
Park, K.B., Lee, J.Y.: New design and comparative analysis of smartwatch
  metaphor-based hand gestures for 3d navigation in mobile virtual reality.
  Multimedia Tools and Applications  \textbf{78}(5),  6211--6231 (2019)

\bibitem{ramos2018new}
Ramos, F., Trilles, S., Torres-Sospedra, J., Perales, F.J.: New trends in using
  augmented reality apps for smart city contexts. ISPRS International Journal
  of Geo-Information  \textbf{7}(12), ~478 (2018)

\bibitem{Roda2014}
Roda, N.: {User centered design process} (2014),
  \url{http://nikkiroda.com/user-centered-design-process/}, accessed on: 9 Jan
  2020

\bibitem{Schinke2010}
Schinke, T., Henze, N., Boll, S.: {Visualization of off-screen objects in
  mobile augmented reality}. In: Proceedings of the 12th International
  Conference on Human Computer Interaction with Mobile Devices and Services.
  pp. 313--316 (2010)

\bibitem{Schneiderman2009}
Schneiderman, B., Plaisant, C.: {Designing the user interface: Strategies for
  effective human-computer interaction}. Pearson (2009)

\bibitem{Spradley2016}
Spradley, J.P.: {Participant observation}. Waveland Press (2016)

\bibitem{Tokusho2009}
Tokusho, Y., Feiner, S.: {Prototyping an outdoor mobile augmented reality
  street view application}. In: Proceedings of ISMAR Workshop on Outdoor Mixed
  and Augmented Reality (2009)

\bibitem{Vredenburg2002}
Vredenburg, K., Mao, J.Y., Smith, P.W., Carey, T.: {A survey of user-centered
  design practice}. In: Proceedings of the SIGCHI Conference on Human Factors
  in Computing Systems. pp. 471--478 (2002)

\bibitem{williams2019smart}
Williams, M., Nurse, J.R.C., Creese, S.: {(Smart) Watch Out! encouraging
  privacy-protective behavior through interactive games}. International Journal
  of Human-Computer Studies  \textbf{132},  121--137 (2019)

\bibitem{williams2019smartwatch}
Williams, M., Nurse, J.R.C., Creese, S.: Smartwatch games: Encouraging
  privacy-protective behaviour in a longitudinal study. Computers in Human
  Behavior  \textbf{99},  38--54 (2019)

\bibitem{Yovcheva2015}
Yovcheva, Z.: {User-centred design of smartphone augmented reality in urban
  tourism context}. {PhD Thesis}, Bournemouth University (2015)

\bibitem{yung2019new}
Yung, R., Khoo-Lattimore, C.: New realities: a systematic literature review on
  virtual reality and augmented reality in tourism research. Current Issues in
  Tourism  \textbf{22}(17),  2056--2081 (2019)

\bibitem{Zhou2008}
Zhou, F., Duh, H.B.L., Billinghurst, M.: {Trends in augmented reality tracking,
  interaction and display: A review of ten years of ISMAR}. In: Proceedings of
  the 7th International Symposium on Mixed and Augmented Reality. pp. 193--202
  (2008)

\bibitem{Zowghi2005}
Zowghi, D., Coulin, C.: {Requirements elicitation: A survey of techniques,
  approaches, and tools}. In: Engineering and Managing Software Requirements,
  pp. 19--46. Springer-Verlag (2005)

\end{thebibliography}

\end{document}